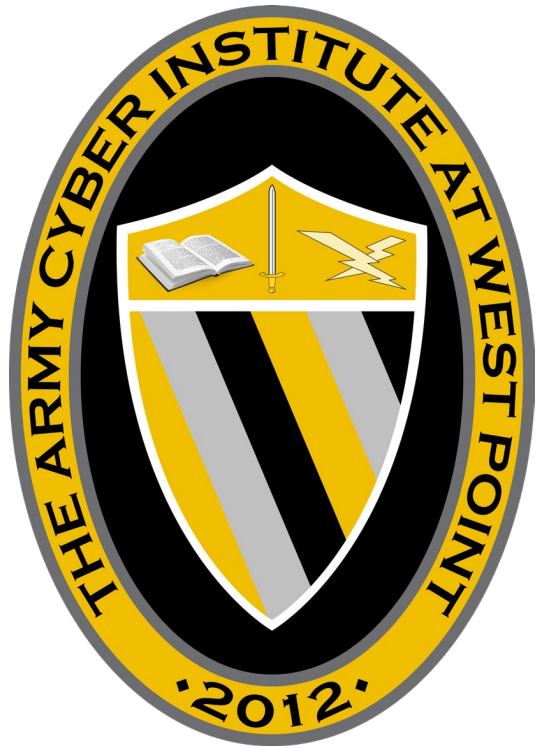

Army Cyber Institute

Technical Report

# Tracking The Trackers: Commercial Surveillance Occurring on U.S. Army Networks

**Date:** May 4, 2025
**Authors:** MAJ Alexander Master, PhD
Jaclyn Fox, PhD
MAJ Nicolas Starck
MAJ Maxwell Love
CPT Benjamin Allison
**Approved by:** Professor Robert Barnsby, Director, ACI
**Contact:** jessica.dawson@westpoint.edu

## Executive Summary

Despite current security implementations, Internet activity on DoD networks is susceptible to web trackers and commercial data collection, which have the potential to expose information about service members and unit operations. This report documents the outcomes of a study to characterize web tracking occurring on Army CONUS unclassified networks. We derived a dataset from the Cloud-Based Internet Isolation (CBII) platform, encompassing data measured over a two-month period in 2024. This dataset comprised the 1,000 most frequently accessed Internet resources, determined by the number of connection requests on CONUS DoDIN-A during the study period. We then compared all domains and subdomains in the dataset against Ghostery's WhoTracks.me, an open-source database of commercial tracking entities. We found that over 21% of the domains accessed during the study period were Internet trackers. The ACI recommends that the Army implement changes to its enterprise networks to limit commercial Internet-based tracking, as well as policy changes towards the same end. With relatively minor configuration changes, CBII can serve as a more effective mitigation against risks posed by commercially available information.

### Key Points

- 21.2% of the top 1,000 Internet resources accessed on Army CONUS unclassified networks during the study period were tracker domains – domain endpoints used exclusively for analytics or collection of user data. This is a conservative estimate of total Internet tracking activity, as another 10.4% consisted of websites with embedded tracking code.

- Of the tracker domains we identified, 26.7% were "advertising." Advertising Technology (AdTech), the code embedded in websites which transmits information to tracker domains, collects a range of sensitive data points (e.g., PII, geolocation, email, home address, personal phone number) from users, aggregates these data into commercially available sources, and has been demonstrated to be exploitable in several recent reports [15, 18, 34, 40].

- Most of the websites visited were common web browsing activities, such as e-learning, Microsoft services, banking, and travel services. The presence of particular domains, including TikTok Analytics, Google China, and a defunct gambling domain, warrants further investigation.

- Communication between users and Internet tracker domains provides no benefit to the Army and introduces risks associated with the collection, aggregation, sale, and sharing of Internet-based commercially available information (CAI).

**Recommendations**

The following recommendations require minimal funding or resources to implement. These actions reflect either incremental changes to the technical implementation of current systems or changes to military policy surrounding these technologies.

1. Ensure tracker domains are not assigned the "allow" policy action in Cloud-Based Internet Isolation (CBII). This will require action from all CBII "mission partners" that operate their own segments of DoDIN-A.

2. Direct that the most commonly used websites that contain tracking code, such as tracking pixels, be added to the "isolate, read-only" policy action in CBII. Reduced reliance on "allow" and "isolate"[1] policy categorizations will ensure that tracking code on websites is removed by CBII before rendering the website and serving it to users.

3. Conduct a review of Intune default configurations for end-user workstation software and group policy objects (GPOs) for NIPR workstations and Azure Virtual Desktop (AVD) to ensure web browser security configurations are specifically configured (by default) to block tracking content, cross-site cookies, and unnecessary scripts or code.[2] These provide protection at the host level and limit commercial data collection from reaching DoDIN network architecture.

4. To increase the efficacy of CBII as a countermeasure for Internet tracking, we recommend adding "Advertising" and "Tracking" to the categories of content in DoDI 8025.AA and other pertinent policies, to ensure that websites engaging in Internet tracking can be easily identified by mission partners and processed by CBII's isolation.

5. Alter contracting language for commercial software used on DoDIN to require the minimization of commercial data collection to only the minimum necessary amount for a system to function, ensure that the government maintains control of any such telemetry or user data, and restrict third-party transfer.

6. Educate service members and civilian employees about the risks posed by commercial tracking, CAI, and data brokers.

---

[1] This implementation alone is sub-optimal when compared to "isolate, read-only" for reducing commercial data collection. If users are logged into authenticated accounts (e.g., Google, Facebook) and a domain is only categorized in the "isolate" policy action, commercial entities will be able to track user activity across devices and sessions server-side, despite the segmentation provided by CBII. The "isolate, read-only" policy action strips the DOM of all unnecessary code, including tracking code, and renders only the static content to the user.

[2] This is necessary given that websites categorized as "allow" are not isolated by CBII; traffic is not containerized in this case, and tracking of users will still occur if host-based protections or DNS blocklisting are not implemented.

# Table of Contents

# Introduction

Individualized information collected during everyday interactions in modern society can be repurposed to target service members and government employees. Businesses continually collect data about their users from a variety of sources and technical methods. This ubiquitous technical surveillance (UTS) [13, 20] is often conducted with the goal of presenting individually tailored advertisements to consumers. Personally identifiable information (PII) is also aggregated into bulk, commercially available information (CAI) datasets by entities such as data brokers [18]. Researchers have shown how CAI can be exploited, with the potential for economic, psychological, and even physical harm to those targeted [15]. Service members also have unique characteristics as an "audience segment" that make them vulnerable to microtargeting [8, 10, 42].

This study investigates the data collected by commercial entities during the routine use of computer systems on the Department of Defense Information Networks (DoDIN) by service members and Department of Defense (DoD) employees. We utilized a dataset containing the top 1,000 Internet resources requested on Continental United States (CONUS) Army unclassified networks during a two-month period in 2024, derived from the Cloud-Based Internet Isolation (CBII) on DODIN-A. We compared the domains with Ghostery's WhoTracks.me, an open-source database of commercial tracking entities [19, 23]. We found that over 21% of the domains accessed during the study period were Internet trackers. We propose recommendations for policy changes, as well as technical configuration changes, that will strengthen CBII's ability to protect service members from commercial data collection.

# Background

Commercial surveillance takes place on military networks in much the same way it does in the private sector. Advertising Technology (AdTech) present in websites collects intrusive kinds of data, such as personally identifiable information (PII), geolocation, email addresses, phone numbers, home address, personal preferences, and more. Commercial websites use these technical measures (e.g., tracking pixels, web beacons, cookies) to collect site analytics,[3] or to follow users across websites and build profiles of their browsing history and activities [23, 41]. Firms known as data brokers [35, 11] often purchase or obtain these data through sharing agreements with companies or intermediaries, aggregating these individualized data points together. The resulting combined datasets can be purchased at low cost, with little to no regulatory oversight, including the ability to purchase identified data about service members and their families [36].

---

[3] Site analytics trackers collect telemetry data (e.g., device identifiers, session identifiers, software versions) and information about how visitors interact with a website, often with the goal of understanding user behavior, measuring engagement, optimizing site performance, or supporting business objectives.

National security vulnerabilities enabled by CAI have also been demonstrated in several recent reports [18, 34, 37]. Journalistic reporting from WIRED demonstrated how commercial data brokerage services expose location data of service members and use it to follow them from their residences to sensitive facilities, such as intelligence organizations or nuclear sites around Europe [29]. Another report outlined how it is possible to target online advertisements to specific audiences throughout the United States (U.S.) federal government, including military service members, executive agency staff of the U.S. Congress, and federal judges [2]. The U.S. government has begun to acknowledge some of the vulnerabilities introduced by CAI, when in 2025, the Department of Justice (DoJ) issued regulatory rules that prohibit the sale or sharing of "sensitive personal data" and "government-related data" to particular foreign countries of concern [39]. Threats of physical harm, such as when Jenn Easterly—then Director of the Cybersecurity and Infrastructure Security Agency (CISA)—was "swatted"[4] at her home in northern Virginia [31] have also motivated rulemaking for the protection of personal information of government officials. An FBI affidavit revealed that the man who shot and killed a Minnesota representative and her husband in their home, along with targeting several other elected state officials, used data broker "people search" websites to find his victims' home addresses [38].

AdTech has also been shown to be a mechanism for malicious cyber actors (MCAs) to deliver malware to victim computers. CISA [5] and the National Security Agency (NSA) [32] have both published guides for federal employees, explaining the security implications of malicious advertising ("malvertising"), whereby MCAs embed malware [26] into payloads delivered by AdTech via a web browser to an individual's computer. A study by Adalytics, observing requests from the top 10,000 websites across the Internet, highlighted that under 30% of the most popular pages loaded ads with sandboxing techniques as a security measure, while the majority did not [1]. While not all AdTech on commercial websites around the Internet is complicit in cybercriminal activity, the potential for this kind of exposure to cybersecurity threats is not an acceptable risk within military networks or government systems in a national security context.

Previous research related to the disclosure of information about government officials and service members has often focused on CAI [15, 16, 36], but it does not explore the commercial data collection occurring on the DoDIN. While the Army has limited authority over many sources of CAI relating to personnel, it does control the Army portion of the DoDIN (DoDIN-A). Understanding the prevalence of CAI collection on DoDIN-A can enable the Army to take informed action to protect military-affiliated personnel who use government networks.

---

[4] Swatting is a form of criminal harassment, used to deceive an emergency service provider into sending a police response team to another person's address due to the false reporting of a law enforcement emergency.

## Methodology

In this study, we analyze the most commonly requested Internet resources on Army CONUS unclassified networks. We obtained a list of the top 1,000 uniform resource locators (URLs) requested on the Army CONUS Non-classified Internet Protocol Router (NIPR) portion of DoDIN-A, from both February and March 2024. These URLs represent a sample of the most common web browsing activities of Soldiers and Army Civilians stationed in the U.S. who have access to DoDIN-A while on duty. Our analysis is based on data from the Cloud-Based Internet Isolation (CBII) program [30], which is utilized across DoDIN NIPR and provided by Menlo Security. We obtained our data from GABRIEL NIMBUS, Army Cyber Command's big data platform.

> **Cloud-Based Internet Isolation (CBII)**
>
> CBII is an Infrastructure as a Service (IaaS) solution intended to isolate browsing behavior associated with cybersecurity risk. CBII provides a "containerized", isolated environment in cloud infrastructure, where a web request is isolated from other browsing activity before being presented to the user's browser. It can also deny access to prohibited content. CBII is intended to prevent cybersecurity threats (e.g., malware) from reaching a user's device, remove trackers and unwanted code from websites, deny access to prohibited content, and reduce throughput requirements on DoD networks. CBII often excludes military and government-owned websites, as well as commercial websites whitelisted[5] by DoD "mission partner" tenants who control governance of their own segment of the DoDIN. U.S. Army Network Enterprise Technology Command (NETCOM) and Joint Force Headquarters (JFHQ) DoDIN ordered the implementation of CBII for all NIPR DoDIN-A users in 2021,[6] with the implementation set to be complete by 2024. Enforcement of DODIN-A CBII policies began in January of 2024.

There are four possible policy action categories within the context of categorization in CBII: "allow", in which the website is not isolated by CBII and there is direct communication between a user and the target website or Internet resource through DoDIN infrastructure; "isolate", in which the website is dynamic/interactive but is isolated in CBII cloud infrastructure; "isolate, read-only", in which the website is rendered non-interactive and is isolated in CBII; and "blocked", in which the website cannot be accessed. Websites in the "allow" category can be accessed directly without any CBII isolation, and third-party trackers are handled by the end user's browser – which allows tracking and collection of their data.

---

[5] A network operator could categorize a commercial website under "allow" in CBII if it does not function properly using CBII isolation. This choice may also introduce vulnerability into the DoDIN.
[6] NETCOM Command Message 2021-08, Major General Maria Barrett

All URLs in the dataset were labeled as being under the "allow" policy action category of the CBII program, meaning that when web requests were made to the domains in question, they bypassed the protections and isolation CBII could provide and traversed DoDIN-A infrastructure normally. This is common practice for ".gov" and ".mil" domains.

We compared the domain/subdomain of each URL against the domains Ghostery's "WhoTracks.me" tracker database to determine if the domain or subdomain was associated with third-party tracking activity. The WhoTracks.Me database is a publicly accessible, empirically derived repository that documents the prevalence and behavior of third-party trackers across the web. The database is maintained through a collaboration between Ghostery and crowdsourced data from academic researchers. The database can be used to identify tracking domains, categorize their functions (e.g., advertising, analytics, social media), and attribute ownership to parent companies.

When querying a target domain, Ghostery's tracker database returns two different results: "websites" and "trackers" for the domain of the URL entered. Tracker domains are the endpoints (e.g., ad.doubleclick.net, analytics.tiktok.com, demdex.net, analytics.[INSERTCOMPANYNAME].com) that receive data from AdTech, such as tracking pixels. Tracking pixels are blocks of code inserted into a website that collect and transmit user data and heuristics [12]. Tracking pixels can be used for a variety of purposes, including: tracking conversions, a marketing term for users that eventually make purchases; retargeting, which includes following users and serving ads based on their preferences to encourage previous customers to make follow-on purchases; and impression tracking, which follows user behavior (e.g., clicks, browsing time, page navigation) to evaluate effectiveness of an ad campaign. Examples include the Meta pixel, Google Analytics, and the TikTok pixel, which are commonly found on many commercial websites.

Ghostery's tracker database may also categorize a (sub)domain as both a website and tracker. For example, WhoTracks.Me returns both values when querying the domain "google.com". Put another way, google.com is a domain individual users can visit where a webpage is rendered, that also contains tracking code. A tracker domain is not typically a website a user can visit, but rather an endpoint for receiving data.

Below we show examples of individual queries using the WhoTracks.Me web interface provided by Ghostery. Figure 1 shows an illustration of a "tracker domain" (omtrdc.net). Clicking on "omtrdc.net" gives more information about the entity that operates the tracker domain, as well as other (sub)domains associated with that entity, as shown in Figure 2. They also provide summary details about "tracker reach", such as how prolific that commercial entity's tracking activity is across the Internet, among other details.

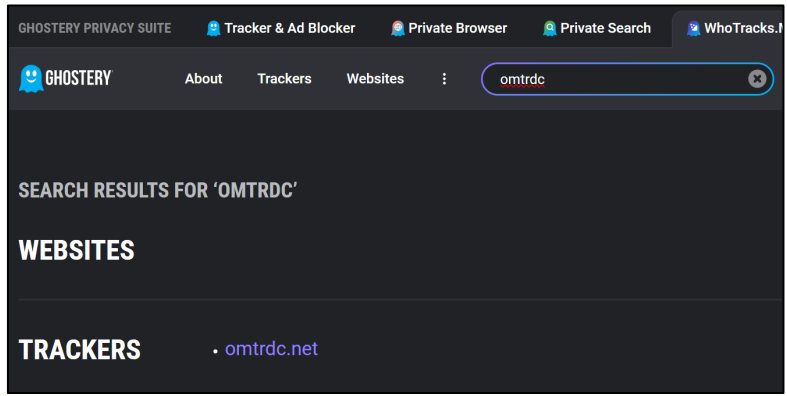

**Figure 1.** Ghostery query, example of a tracker domain

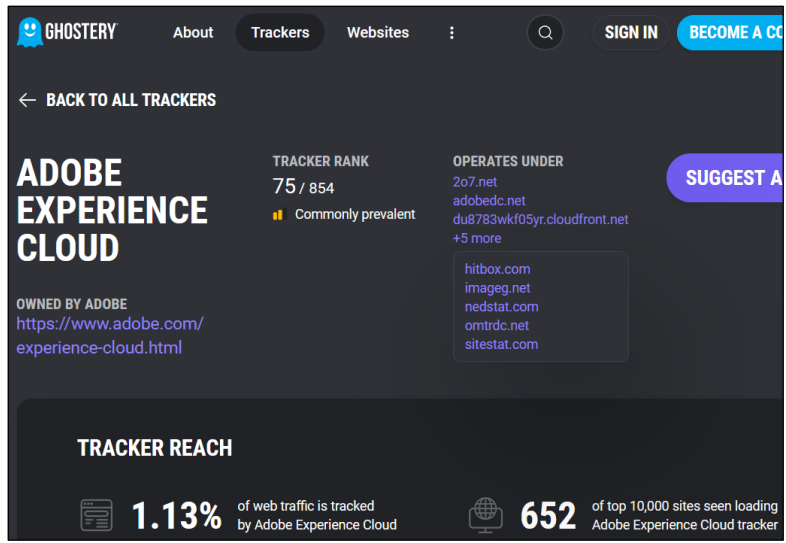

**Figure 2.** Ghostery query, example of a tracking entity and its tracker domains

Ghostery may include results in both categories (websites and trackers). For example, the domain "google.com" returns the result of being both a website and a tracker, as shown in Figure 3. We can conclude, based on Ghostery's methodology [16], that google.com is a domain individual users can visit where a webpage is rendered, that also contains tracking code.

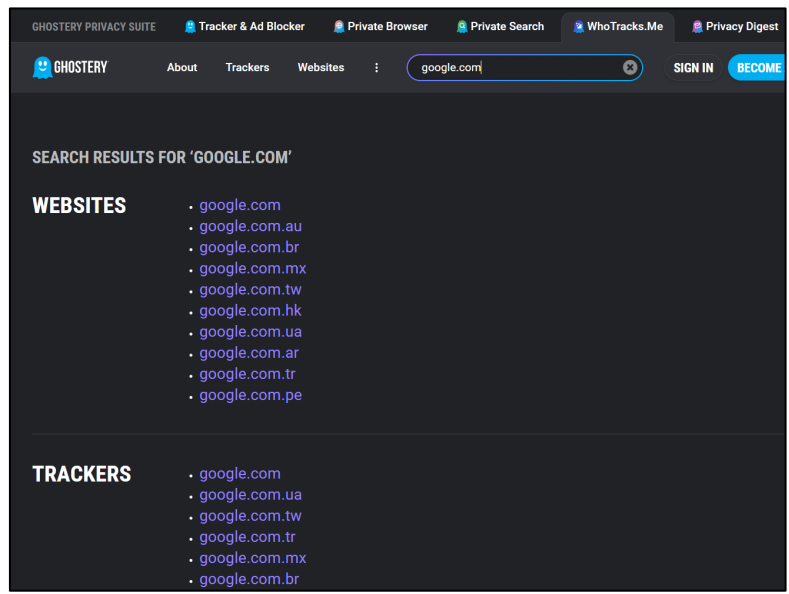

**Figure 3.** Ghostery query, example of a website categorized for tracking activity

Given the scope of our study, the analysis that follows focuses primarily "tracker domains" (e.g., omtrdc.net). These domain endpoints are specifically designed to track user behavior. However, a limitation of this view is that it provides a conservative estimate of overall tracking activity, given that AdTech is prolific and present on many websites that are not exclusively trackers themselves. Figure 3 illustrates an example of a front-facing website that includes tracking code, but is not specifically a tracker domain.

**Data Availability:** The raw datasets used in this report will not be made available to the public. However, DoD personnel with appropriate access and need to know who would like to see the data or replicate our study are encouraged to contact the authors.

In the next section, we report our statistical analysis of total tracking activity, categories of tracker capability (e.g., site analytics, advertising), and entities/organizations that operate tracker domains (or websites with tracking ability).

# Findings

## Summary Statistics

Of the 1,000 websites in our analysis, 316 are found in the Ghostery database as either tracker domains (212) or websites with tracking ability (104). See Table 1. Tracker domains comprise over **21%** of the top 1,000 Internet resources accessed on Army CONUS networks during the study period.

**Table 1:** Domain Designations in Ghostery

| Domain Designation | Count | Percentage |
|---|---|---|
| Tracker Only | 212 | 21.2 |
| Website with Tracking Ability | 104 | 10.4 |
| Domain without Tracking, or Unlisted | 684 | 68.4 |

N = 1,000

## Categories of Trackers

Ghostery provides contextual categorization when comparing a domain or URL with their WhoTracks.me database. Analysis of the tracker domains by category reveals that nearly half engage in activities related to site analytics. More than one-quarter, however, are explicitly advertising-related. This indicates that 57 of the top Internet resources that made connections on DoDIN-A CONUS networks (as represented in our dataset) are trackers related to advertising activities. Figure 4 shows a summary by percentage of the tracker categories for the tracker domains.

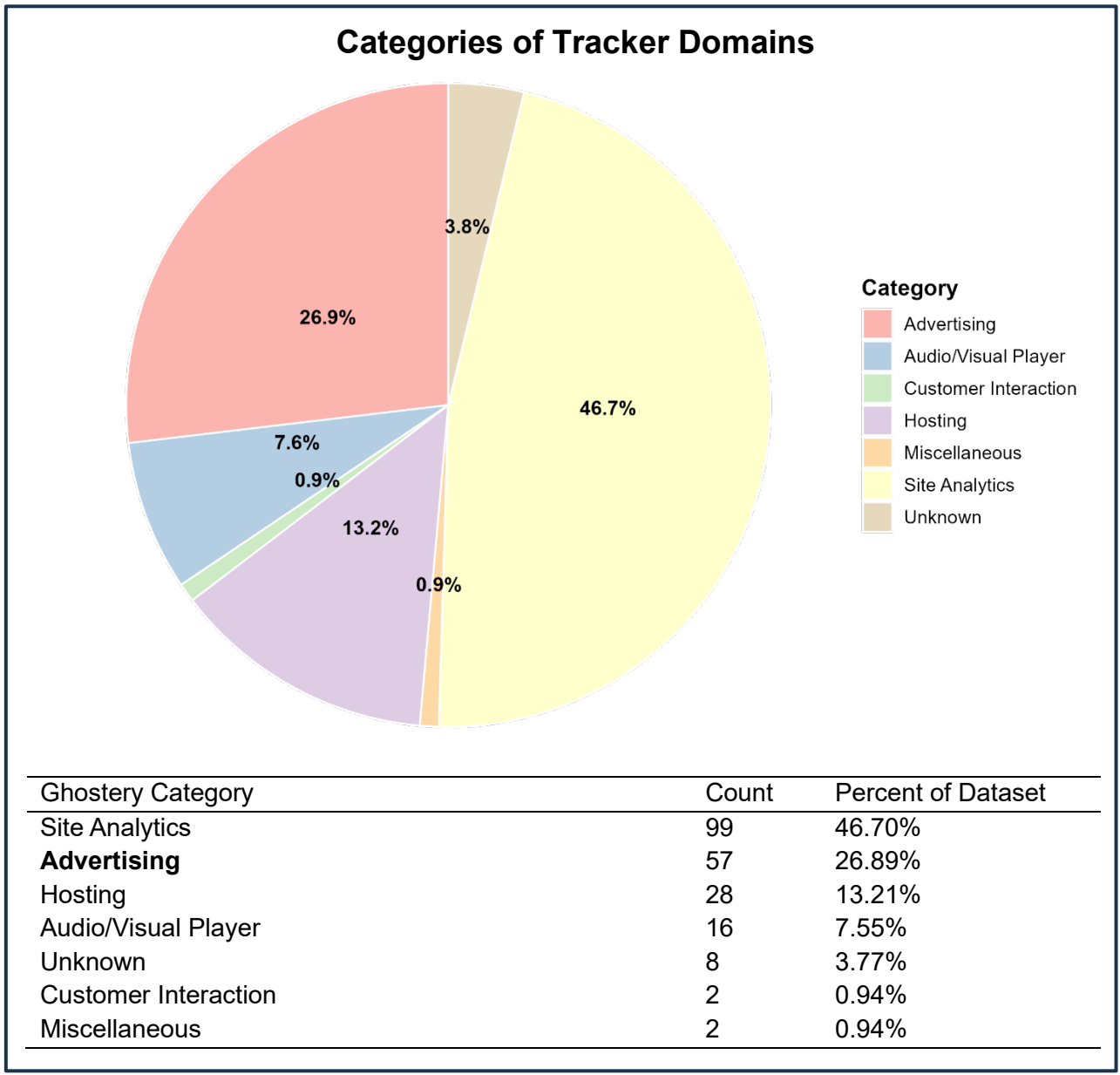

**Figure 4.** Categories of Tracker Domains

Next, we map the Ghostery categories for all websites with tracking ability, as shown in Figure 5. That is, the tracker domains as well as the domains that are both websites and have tracking ability. On websites with embedded trackers, site analytics remains the largest category. Advertising domains are the second largest category after site analytics. The increase in advertising proportion is largely due to Google domains, which are categorized as advertising by Ghostery's database.

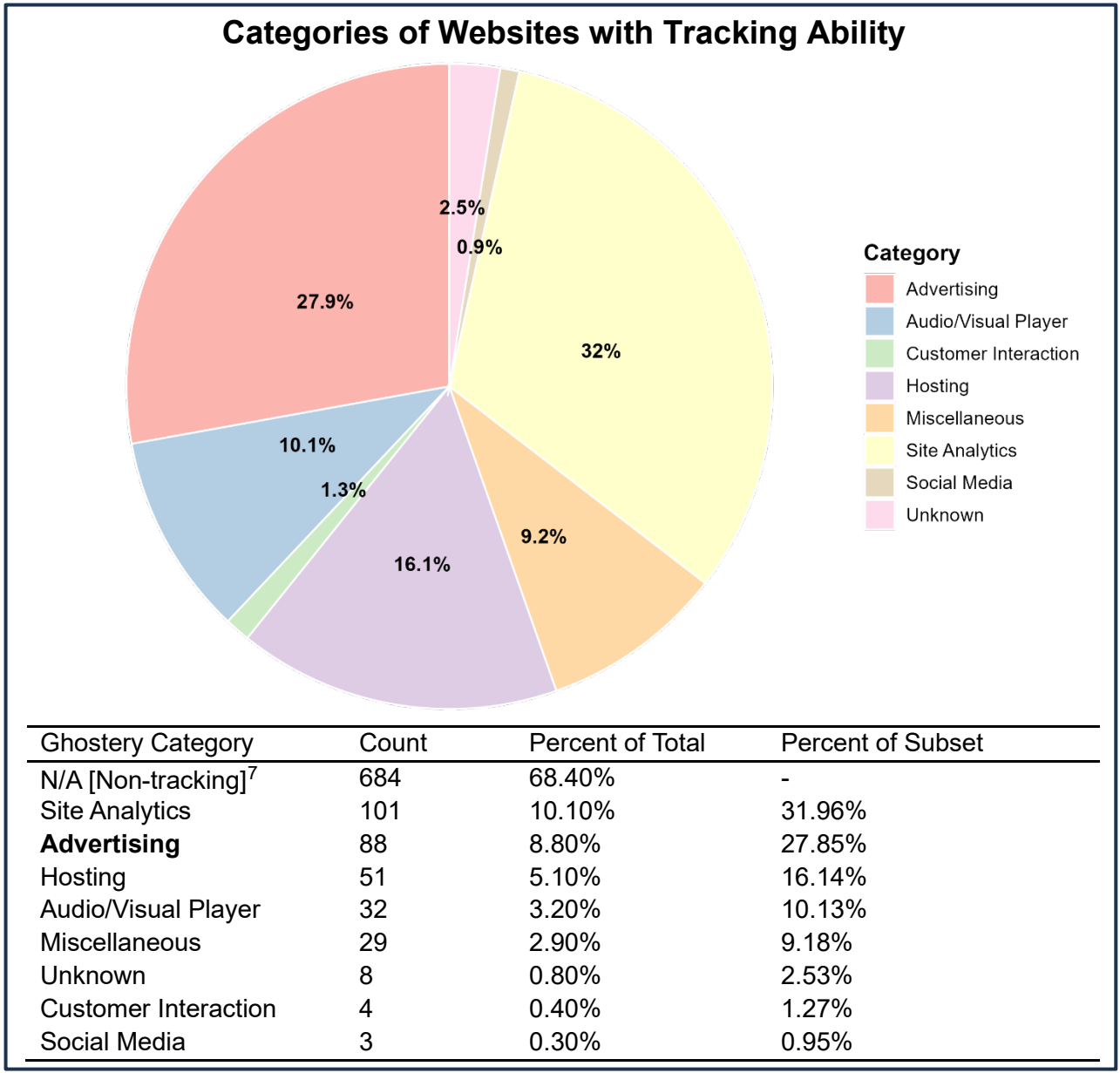

**Figure 5.** Categories of Websites with Tracking Ability

---

[7] "Non-tracking" in these tables refers to domains that were not categorized by Ghostery's WhoTracks.me database as either tracker domains (see Figure 1) or as websites that include tracking code (see Figure 3) at the time our study was conducted. In reality, some of these websites may have trackers that have not been categorized – or may have added tracking code since 2024.

**Tracker Domains Analysis**

In Table 2, we highlight the AdTech companies that are associated with tracker domains.[8] In our dataset, Adobe is the most frequent actor, operating many different tracker domain endpoints. The top two entities, Adobe Experience Cloud and Marketo (owned by Adobe Inc.), together account for over 36% of the total tracker domains in our sample. It is unclear whether the DoD is knowingly allowing Adobe and Microsoft, both software providers to the DoD, to track user behavior, or if the tracking occurs by default in the provided software (as is common for private sector commercial customers). Other top entities appearing in our dataset include Akamai Technologies, Quantum Metric, Contentsquare, and Datadog. Notably, TikTok (owned by the Chinese company ByteDance) operates one of the tracker domains on the list, analytics.tiktok.com. Additionally, Criteo, a company headquartered in France and known for its cross-device tracking capabilities, appears in our data. Criteo could potentially link users of government devices to their personal devices to reveal patterns of life or other details detrimental to force protection [7]. Table 2 is sorted alphabetically.

**Table 2:** Entities That Operate Tracker Domain Endpoints

| Entity | Category | Number of tracker domain endpoints in the top 1000 dataset |
|---|---|---|
| Adobe Audience Manager | Advertising | 1 |
| Adobe Experience Cloud | Site Analytics | 49 |
| Ads Ninja | Advertising | 1 |
| Akamai Technologies | Hosting | 27 |
| Appnexus | Advertising | 1 |
| Audigent | Advertising | 1 |
| Bidswitch | Advertising | 1 |
| Bugsnag | Site Analytics | 1 |
| Clicktale | Site Analytics | 2 |
| Connatix.com | Advertising | 2 |
| Contentsquare.net | Advertising | 5 |
| Criteo | Advertising | 1 |
| Dailymotion | Miscellaneous | 1 |
| Datadog | Unknown | 5 |
| Dotomi | Advertising | 1 |
| Doubleclick | Advertising | 1 |
| Facebook CDN | Hosting | 1 |
| Foresee | Unknown | 1 |
| Forter | Unknown | 1 |
| Google Signals | Advertising | 1 |
| Google | Customer Interaction | 1 |
| Gullstory | Site Analytics | 1 |
| Index Exchange | Advertising | 1 |
| Inmobi | Site Analytics | 1 |
| JW Player | Audio/Video Player | 2 |
| Launch Darkly | Unknown | 1 |

---

[8] Tracker domains, regardless of which company operates them, are of concern because their purpose is to gather user information and telemetry. Once collected, it is difficult to determine how that data will be used or if it will be sold or shared with other entities.

| Entity | Category | Number of tracker domain endpoints in the top 1000 dataset |
|---|---|---|
| Marketo (owned by Adobe Inc.) | Advertising | 28 |
| Media.net | Advertising | 1 |
| Microsoft Clarity | Site Analytics | 24 |
| MSN | Miscellaneous | 1 |
| Mux Inc. | Audio/Video Player | 11 |
| Namogoo | Advertising | 1 |
| New Relic | Site Analytics | 4 |
| Openx | Advertising | 1 |
| Pendo.io | Site Analytics | 1 |
| Permutive | Advertising | 1 |
| Pub.network | Advertising | 1 |
| Pubmatic | Advertising | 1 |
| Quantum Metric | Site Analytics | 15 |
| Rubicon | Advertising | 2 |
| Salesforce Live Agent | Customer Interaction | 1 |
| Soundcloud | Audio/Video Player | 1 |
| Sovrn | Advertising | 1 |
| Tiktok Analytics | Site Analytics | 1 |
| Tradedesk | Advertising | 1 |
| Triplelift | Advertising | 1 |
| Unruly Media | Advertising | 1 |
| Vimeo | Audio/Video Player | 2 |

## Discussion

As shown in the results above, **21.2% of the top Internet resources accessed on the network were tracker domains**. This percentage is a conservative estimate of the total amount of tracking activity, given that Ghostery separately categorizes front-facing websites that include tracking code but are not specifically tracker domains.

The majority of trackers we identified were categorized as site analytics (46.7%). Site analytics include telemetry about a user's browsing experience, such as number of clicks, time spent on pages, web browser characteristics, and geolocation of a user based on IP address. Individually, these data points may be benign. However, when combined with browser fingerprinting [28] methods (which are increasingly implemented on e-commerce websites) and aggregated into CAI data sets, they could potentially be used to identify individual users and track them across devices.

Most concerning is the next most frequent category of tracker domains, those used for advertising purposes (26.9%). AdTech (e.g., tracking pixels, cookies, browser fingerprinting) collects individually identifiable data, such as name, geolocation, email, phone number, home address, and more. Personally identifiable data derived from (and aggregated into) commercial sources has been demonstrated to be exploitable – with the potential for economic, psychological, and even physical harm to those targeted [15]. Service members, military family members, and veterans have unique characteristics as an "audience segment" that make them vulnerable to microtargeting [8, 10, 42].

Our study also illustrates the most common commercial entities present in the dataset. This commercial tracking is not unique to DoDIN-A, as demonstrated in previous research studies [33, 41]. The most frequent organization identified as operating tracker domains in unclassified CONUS traffic in our dataset is Adobe. This finding is unsurprising, given the ubiquity of Adobe's tracking technologies across the broader Internet and the fact that many websites leverage Adobe products (embed Adobe trackers) to gain insight about their customers. The military services also purchase Adobe software (e.g., Adobe Acrobat), which may also communicate with tracker domains. It is unclear whether this tracking is deliberately allowed on DoDIN-A or if it has not been addressed. Software often functions normally even if traffic destined for analytics and telemetry endpoints is blocked; future research should analyze the impacts of large-scale blocking of tracker domains to assess their impacts (if any) on commercial applications and webpages on DoDIN-A. It is important to note that although Adobe operates more tracker domains than others, this is not necessarily indicative of more invasive tracking or higher throughput of data requests, which would require further investigation and a richer dataset to answer.

Overall, the most common Internet resources accessed on CONUS networks tended to involve everyday activities such as finance (e.g., USAA, Citibank, Fidelity, Capital One), Microsoft services, online education (e.g., University of Maryland, WGU, Liberty University), and travel booking services.

Some domains present in our study's dataset may warrant increased examination or scrutiny, such as Google China ("www.google.cn"), the analytic endpoint domain for connections from the TikTok pixel ("analytics.tiktok.com"), and a website allegedly for gambling ("www.grlvelxstuff.com") which closely mimics [24] the domain name of a website promoting weather analysis software ("www.grlevelx.com").

It is unclear if the potentially problematic domains were accessed by individuals or were part of an automated piece of software (or malware). Chinese companies also have drop-in replacements for many Western websites that serve users in mainland China (Baidu, Weibo, WeChat, etc.). Only Chinese-speaking individuals outside of China, expatriates, or international people studying Chinese likely benefit from Google China's website. In the Army, open-source intelligence (OSINT) specialists are generally not permitted to use their administrative NIPR computers to perform operational research, and those queries should not appear in DoDIN-A CBII data, as was used in our study.

The operators of the gambling website that was included in our top 1,000 list appear to have let their transport layer security (TLS) certificates expire in May 2024. This triggers a full-screen red warning in most web browsers when someone navigates to the page, and would also cause most automated requests to the webpage to fail due to encryption errors from a lack of valid certificates. This suggests that the domain is no longer maintained. It is unclear why the domain was accessed frequently enough to make the top 1,000 list.

Finally, we highlight the domain endpoint for the TikTok Pixel. The TikTok pixel is a piece of tracking code inserted into a website to allow for the collection of user data and analytics; its domain endpoint (analytics.tiktok.com), where the pixel transmits data, was in our top 1,000 dataset. This is potentially problematic, given that in 2020, POTUS issued an executive order prohibiting all transactions between anyone under the jurisdiction of the U.S. federal government and ByteDance, the parent company of social media platform TikTok.[9] In December 2022, the *No TikTok on Government Devices Act* was signed into law. Subsequent policy in the executive branch has restricted the use of TikTok's applications on government devices, given concerns about how corporations under Chinese law are obligated to cooperate with the intelligence services of the PRC [6].  Although private companies in the U.S. are not subject to all of the same policies as federal agencies, users of websites that utilize the TikTok pixel are exposed to some of the same risks in terms of commercial data collection.

Communication between government-owned devices and Internet tracker domains provides no benefit to the Army, consumes throughput resources, and introduces risks associated with the collection, aggregation, sale, and sharing of Internet-based commercial data. Army Red Team assessments, ACI reports, and academic studies demonstrate the force protection and operations security (OPSEC) implications of data harvested by commercial entities, such as PII from AdTech and information derived from Internet-based engagement. The ACI has illustrated many of the vulnerabilities introduced by commercial data through research efforts of its Digital Force Protection team [16]. The Threat Systems Management Office (TSMO) has demonstrated Army unit OPSEC failures based on publicly available information (PAI) gathered from social media and other public sources online. Other academic studies have demonstrated how proxy variables in CAI can be utilized to identify individuals with insider access and predispositions for engaging in threat actions against their organization [15]. The risks associated with CAI collection on Army personnel and missions [18] warrant deliberate consideration of available means to enhance the force's operational and force protection posture in cyberspace.

---

[9] Executive Order 13942 (Addressing the Threat Posed by TikTok, and Taking Additional Steps To Address the National Emergency With Respect to the Information and Communications Technology and Services Supply Chain), which was revised under Executive Order 14034 (Protecting Americans' Sensitive Data from Foreign Adversaries)

## Recommendations

The ACI recommends that the Army implement the following changes across its enterprise networks to limit commercial Internet-based tracking; doing so will further mitigate risks to service members and unit readiness posed by commercial data. Employing measures to block tracker domains and cross-site analytics will help to limit the collection of data on service members, reduce organizational footprints in CAI datasets, and reduce bandwidth requirements by blocking unnecessary traffic. Army Cyber Command (ARCYBER), DoD Cyber Defense Command (DCDC), and CBII mission partners across the service component already possess the relevant authorities to promulgate policy for blocking and filtering connections and content on DoDIN-A and have the capabilities to implement these policies without extra resources.

NETCOM and other mission partners should ensure that tracker domains are not assigned the "allow" policy action in CBII. ARCYBER should direct all mission partners to add tracker domains and websites that contain tracking code (tracking pixels, web beacons, browser fingerprinting) to the "isolate, read-only" policy action in CBII. To take mitigations a step further, they could direct that all uncategorized websites be designated as "isolated, read-only" to ensure protections remain as future tracking domains are created and gain prominence. Reduced reliance on "allow" and "isolate"[10] policy categorizations will ensure that tracking code on websites is removed by CBII before rendering websites and serving them to users.

ARCYBER should direct a review of Intune default configurations for end-user workstation software and group policy objects (GPOs) for NIPR workstations and Azure Virtual Desktop (AVD) to ensure web browser security configurations (e.g., Microsoft Edge, Mozilla Firefox, Google Chrome[11]) are specifically configured to block tracking content, cross-site cookies, and unnecessary scripts or code.[12] This defense-in-depth approach of stopping tracker code from executing in a user's browser (as well as blocking tracker domains at the network edge[13] or isolating them with CBII) will mitigate much of the risk posed by trackers before they reach DoDIN infrastructure. Browsers should also be configured to transmit Global Privacy Control (GPC) and Do Not Track (DNT) signals by default. For commercial websites that honor these signals,[14] this would automatically opt-out the user from commercial data collection.

---

[10] The "isolate" implementation alone is sub-optimal when compared to "isolate, read-only," which strips the Document Object Model (DOM) of the webpage of all executable code, and renders solely the page content to the user. If users are logged into authenticated accounts (e.g., Google, Facebook) and a domain is only categorized in the "isolate" policy action, commercial entities will be able to track user activity across devices and sessions server-side, despite the segmentation provided by CBII.

[11] Notably, Google Chrome is the only major web browser that has not implemented blocking of third-party cookies. Google initially planned to follow other web browsers in implementing the blocking of third-party cookies to promote user privacy and limit cross-device tracking, but abandoned these plans in July 2024 [3]. Given the privacy risks posed by Google Chrome as a browser [27], we recommend that Army network maintainers restrict the installation of Chrome on NIPR workstations and AVD.

[12] This is necessary given that websites categorized as "allow" are not isolated by CBII.

[13] Domain Name System (DNS) blocklisting of tracker domains complements all of these approaches.

To increase the efficacy of CBII as a countermeasure for Internet tracking, we recommend adding "Advertising" and "Tracking" to the content categories in DoDI 8025.AA and other pertinent policies, to ensure websites engaging in Internet tracking can be easily identified by mission partners and processed by CBII's isolation. Army CIO/G-6 and other relevant stakeholders could socialize these changes to the DoD.

The Army and DoD can also engage with commercial entities to open a dialogue about reducing their commercial data collection practices. Prior research has shown that vendor notification is an effective means to effect business change decisions; one academic study in 2022 revealed Meta pixel usage across healthcare patient portals online [22]. The paper resulted in a pervasive electronic health record (EHR) vendor changing their technical implementations, along with several individual hospital systems removing the Meta pixel from their websites to avoid "bad press" and ensure HIPAA[15] compliance. The DoD and military services should also alter contracting language to require the minimization of commercial data collection to only the bare minimum necessary for a system to function, and ensure that the government maintains control of any such data. Adding clauses that restrict third-party transfer of data would provide the government with further assurance that its information will not be misused.

## Limitations and Future Research

This report highlights only a snapshot in time of Army CONUS web traffic. DoD network configurations change, and commercial websites may add or remove AdTech from their websites over time.

The vantage point of our dataset is only from the perspective of CBII and is not end-to-end. It is possible that some of the tracker domains found in the CBII "allow" category are being blocked by their IP addresses or "DNS sinkholed" further up in the DoDIN stacks near the Internet Access Points (IAPs) or beyond. Even if this is the case, we argue that stopping tracking closer to the user, via web browser configurations or CBII, provides more effective "defense-in-depth" protection from commercial data vulnerabilities.

---

[14] Commercial entities under the jurisdiction of the General Data Protection Regulation (GDPR) in the European Union, the California Consumer Privacy Act (CCPA) in the U.S., or other state privacy laws, are often obligated to respect the privacy opt-out choices from cookie banners or presence of a GPC signal from a web browser.

[15] The Health Insurance Portability and Accountability Act (HIPAA) is a federal law that establishes national standards for protecting the privacy and security of individually identifiable health information, also known as protected health information (PHI). The law applies to healthcare providers, health plans, and healthcare clearinghouses that conduct electronic transactions, mandating the protection of patient information.

CBII cannot prevent all commercial tracking activity on its own. Telemetry collection embedded into desktop software (e.g., Adobe products, Microsoft Office) contacts tracker endpoints outside the context of a web browser – and thus CBII cannot provide isolation or tracking reduction for natively installed applications. However, the ubiquity of AdTech on websites, coupled with increasing reliance on web apps and services provided in browsers, makes CBII a highly impactful privacy and security control measure.

Future research should seek to quantify the impact of implementing the recommendations in this report. Comparing the total quantity of tracker domains tagged as "allow" in a future study would validate that CBII is isolating and removing tracking code from more user-requested webpages than it is currently.

Future research should also investigate the commercial tracking activity occurring on government-owned devices. Devices such as smartphones utilize commercial cellular providers to connect to Internet resources, and do not benefit from DoDIN security implementations or CBII. Further investigation into whether contracting language explains appropriate data usage to software providers who are collecting and selling data obtained from government devices is also warranted.

With the introduction of bring-your-own-device (BYOD) policies and programs, Soldiers and Army Civilians connect their personal devices to the Army's information technology (IT) infrastructure. While some mitigations are in place to protect government data (e.g., Azure Virtual Desktop (AVD), Hypori Halo for mobile) to ensure Army data does not reside on end-user devices, third-party trackers and application APIs/SDKs still collect sensitive information about personnel from their computers/smartphones and transmit them over whatever connection Soldiers are using (e.g., home Wi-Fi, cellular networks). Future studies should assess the kinds of information being collected by commercial applications targeted toward service members, military families, and veterans for use, particularly on personal devices.

## Conclusion

This study illustrates that commercial tracking occurs on Army unclassified computer networks. Over 21% of the top 1,000 domains accessed from CONUS DoDIN-A within the two-month period of our study were Internet trackers. This assessment is conservative based on our methodology, as many websites do have embedded tracking code, but are not themselves tracking domains. Because Internet tracking can threaten force protection, OPSEC, and cybersecurity of the force, these findings warrant further action. We offer recommendations for the Army to mitigate risks posed by the collection, sale, aggregation, and sharing of commercial data from DoDIN-A.

# Appendix

## Second Round of Data Collection

After completing the initial round of data collection and analysis, we collected additional data for comparison and replication purposes. The second dataset included the top 1,000 (sub)domains on DoDIN-A CONUS unclassified networks from November and December 2024. Our first objective was to gauge whether the baseline of accessed Internet resources looked relatively consistent over time. Secondly, we wanted to determine if the quantity or diversity of tracker domains had changed, which could inform or alter our recommendations.

The sections below describe the findings from our second iteration of the study. Overall, many of the domains in the top 1,000 matched the first dataset, suggesting that, in aggregate, the baseline of the kinds of traffic on DoDIN-A is similar over time. There was a proportionally small decrease in the amount of tracker domains requested, 19.2% in November and December 2024, rather than the 21.2% observed from February and March 2024. Site analytics remained the most prevalent among the trackers, while advertising (the most problematic category) remained the second highest. All of the "domains of concern" (TikTok Analytics, Google China, grxlevelstuff) from early in 2024 appeared in our second dataset, suggesting no changes had been implemented to restrict their access or put them into CBII isolation. As one example of scale, the TikTok pixel had 180,009 connection requests (#174 out of 1000) while the lowest domain in the dataset (#1000, learning-experience-api-gateway.wgu.edu) had 2122 connection requests.

## Summary Statistics

Of the 1,000 URLs in our analysis, 290 are found in the Ghostery database as either tracker domains (192) or websites with tracking ability (98). See Table 3. In the November-December pull, tracker domains accounted for over 19% of the top 1,000 requests. This is less than the 21% found in dataset 1 for February and March 2024.

In addition to replicating our earlier analyses, the second data pull enabled us to access an additional variable that captured the number of connections to each of the top 1,000 URLs. This is significant as the top 1,000 URLs are not uniform in their frequency of access; rather, the most visited sites have significantly more hits than the least. By examining this variable, we gain a deeper understanding of how prominent tracker domains actually are in our data. To assess this, we summed the counts of visits for each of the top 1,000 URLs, resulting in 2,285,724,454 hits. We then counted the number of visits for the tracker domains only, which gave us 957,499,496. This means that approximately 41.89% of the total visits for the top 1,000 URLs involved tracker domains. That is, <u>tracker domains made up 19.2% of the top domains in the dataset, but 41.89% of the total web requests.</u>

Unfortunately, we were unable to retroactively access this variable for the first dataset (February to March 2024) due to data retention policies, so we are unable to compare connection request counts across time between the two study segments. However, this provides us with additional evidence regarding the ubiquity of tracking on CONUS DoDIN-A unclassified networks.

**Table 3:** Domain Designations in Ghostery

| Domain Designation | n | Percent |
|---|---|---|
| Tracker Only | 192 | 19.2 |
| Website with Tracking Ability | 98 | 9.8 |
| Domain without Tracking, or Unlisted | 710 | 71.0 |

N = 1,000

**Categories of Trackers**

When comparing the first and second data pulls with regard to the Ghostery Categories for tracking activity, we find that the top four categories (site analytics, advertising, hosting, and audio/video player) remained in the same order of most to least prevalent, with only minor deviations in raw numbers and percentages. However, the second pull also included trackers with additional categories, such as utilities (n=10), consent management, social media, and tracking, each of which had one associated domain. There were also more customer interaction domains in the later pull and fewer "unknowns". Of note, "unknown" represented an error with the Ghostery database query (i.e., missing data) as opposed to a defined category in and of itself.

**Table 4:** Categories of Tracker Domains

| Ghostery Category | Count | Percent |
|---|---|---|
| Site Analytics | 92 | 47.92% |
| Advertising | 46 | 23.96% |
| Hosting | 22 | 11.46% |
| Audio/Video Player | 10 | 5.21% |
| Utilities | 10 | 5.21% |
| Customer Interaction | 6 | 3.13% |
| Misc | 2 | 1.04% |
| Consent Management | 1 | 0.52% |
| Social Media | 1 | 0.52% |
| Tracking | 1 | 0.52% |
| Unknown | 1 | 0.52% |

When comparing the February-March and November-December data pulls for the expanded category of tracker domains and domains for forward-facing websites with tracking ability, we find that the top two categories remain site analytics and advertising, respectively. The raw numbers are also fairly consistent across runs. Notably, hosting accounts for a smaller proportion of the second data pull, with 26 domains compared to 51 in the first pull. This shift moved it from third to fourth place in our follow-up. Audio/Visual and miscellaneous categories were similar across both datasets. However, the "utilities" category was present for the first time in the later sample, along with the consent management and tracking categories.

**Table 5:** Categories of Websites with Tracking Ability

| Ghostery Category | Count | Percent of Total | Percent of Subset |
|---|---|---|---|
| NA [Non-Tracking][16] | 710 | 71.00% | |
| Site Analytics | 103 | 10.30% | 35.52% |
| Advertising | 86 | 8.60% | 29.66% |
| Audio/Video Player | 28 | 2.80% | 9.66% |
| Hosting | 26 | 2.60% | 8.97% |
| Misc | 24 | 2.40% | 8.28% |
| Utilities | 10 | 1.00% | 3.45% |
| Customer Interaction | 7 | 0.70% | 2.41% |
| Social Media | 2 | 0.20% | 0.69% |
| Unknown | 2 | 0.20% | 0.69% |
| Consent Management | 1 | 0.10% | 0.34% |
| Tracking | 1 | 0.10% | 0.34% |

---

[16] "Non-tracking" in these tables refers to domains that were not categorized by Ghostery's WhoTracks.me database as tracker domains (see Figure 1) or as websites that include tracking code (see Figure 3) at the time our study was conducted. In reality, some of these websites may have trackers that have not been categorized – or may have added tracking code since 2024.